\documentclass{naturxiv}
\usepackage[pdftex]{graphicx}
\usepackage[english]{babel}  %
\usepackage{amssymb}         %
\usepackage[fleqn]{amsmath}         %
\usepackage{enumerate}
\usepackage{cite}
\usepackage[T1]{fontenc}
\usepackage{makeidx}
\usepackage[normalem]{ulem}
\usepackage{color}
\usepackage{bm}%
\usepackage{hyperref}   %
\usepackage{booktabs} %
\usepackage{tabularx} %
\usepackage{wrapfig}
\usepackage{pdfpages}
\title{Nanoscale artificial intelligence: creating artificial neural networks using autocatalytic reactions}
\author{Filippo Simini$^{1}$}
\begin{document}
\maketitle
\begin{small}
\begin{affiliations}
 \item Department of Engineering Mathematics, University of Bristol, Merchant Venturers Building, Woodland Road, BS8 1UB, Bristol, UK.
\end{affiliations}
\end{small}
\begin{abstract}

A general methodology is proposed to engineer a system of interacting components (particles) which is able to 
self-regulate their concentrations in order to produce any prescribed output  
in response to a particular input. 
The methodology is based on the mathematical equivalence between artificial neurons in neural networks and species in autocatalytic reactions, 
and it specifies the relationship between the artificial neural network's parameters and the rate coefficients of the reactions between particle species. 
Such systems are characterised by a high degree of robustness as they are able to reach the desired output despite disturbances and perturbations of the concentrations of the various species.

\end{abstract}

\begin{figure}
\begin{center}
\includegraphics[type=pdf,ext=.pdf,read=.pdf,width=175mm]{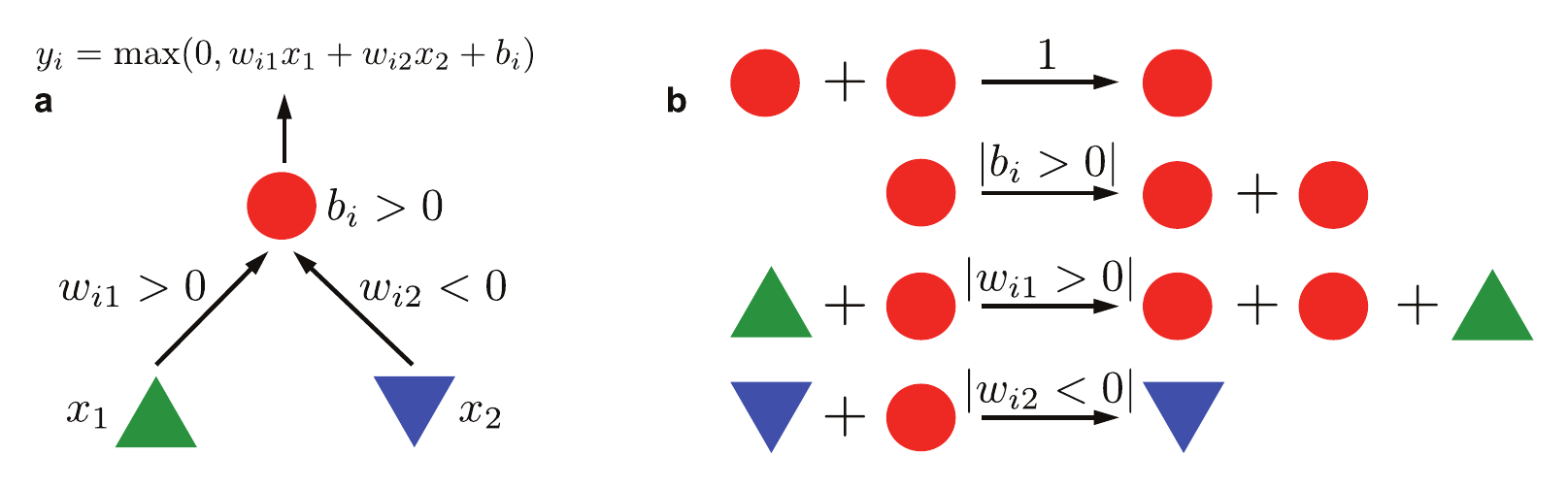}
\end{center}
\caption{{
Equivalence between artificial neurons and autocatalytic reactions. 
a) An artificial neuron (red circle) with two inputs (green and blue triangles).  
The neuron takes as input the values $x_1$ and $x_2$, produces a linear combination using the weights $w_{i1}$, $w_{i2}$ and the bias $b_i$, and finally uses the rectifier activation function, $a(z)=\max(0,z)$, to compute the output $y_i = \max(0, w_{i1} x_1 + w_{i2} x_2 + b_i )$. 
b) The equivalent set of reactions. The artificial neuron and its inputs correspond to different species of particles, represented here as a red circle, and blue and green triangles. 
The output and input values correspond to the concentrations of particles of the three species: $y_i$ is the concentration of red circles, and $x_1$, $x_2$ are the concentrations of green and blue triangles. 
The parameters of the artificial neural network are related to the rates of the reactions between the three species of particles: for example, the weight $w_{1i}$ is proportional to the probability that when a particle of species red circle meets a particle of species green triangle they interact to produce an additional particle belonging to the red circle species. 
A positive weight or bias determines the production of an additional particle of the output species, red circle in this case, while a negative weight or bias determines the elimination of one particle of the output species. 
} \label{fig:1}}
\end{figure}

\begin{figure}
\begin{center}
\includegraphics[type=pdf,ext=.pdf,read=.pdf,width=175mm]{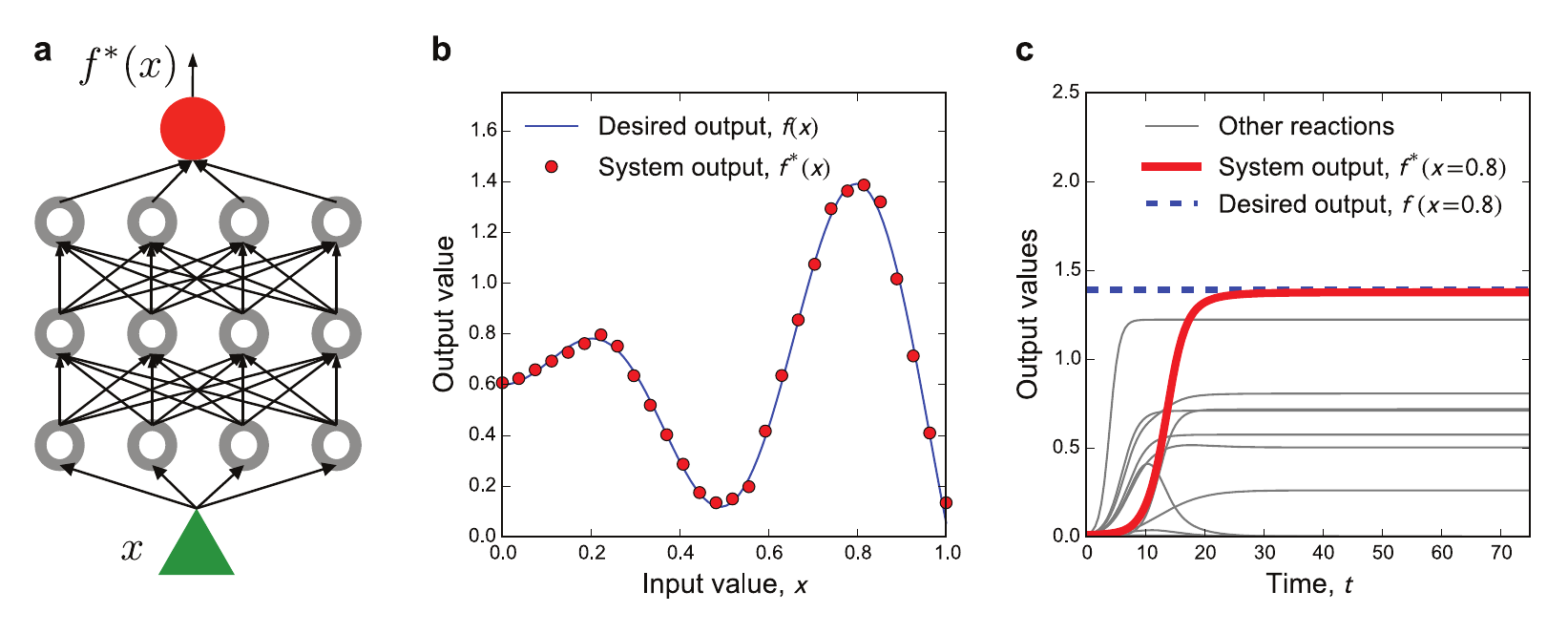}
\end{center}
\caption{{
a) An artificial neural network composed of three hidden layers with four neurons per layer (grey empty circles) and an output layer with one neuron (red circle). 
All neurons use the rectifier activation function. 
The neurons of the first (bottom) hidden layer receive the input value $x$, process it as described in Fig.~\ref{fig:1}a, and produce an output value that the neurons of the next layer use as input. 
b) The network displayed in panel a) has been trained to approximate a desired output, which is the function $f(x)  = x \sin(10 x) + 0.6 $, indicated with a blue solid line. 
The red circles indicate the output of the neural network, i.e. the system output $f^{\ast}(x)$ which is the output value of the neuron in the last (top) layer.
c) A system of species and reactions equivalent to the artificial neural network is created as described in the main text and Fig.~\ref{fig:1}. 
The deterministic rate equation \ref{eq:req} is integrated to obtain the time evolution of the concentrations of all species when the input value $x$ is equal to $0.8$. 
The initial concentrations are $0.01$ for all species. 
Grey lines indicate the concentrations of the species corresponding to neurons in the hidden layers (the grey empty circles of panel a), while the red line indicates the concentration of the output species corresponding to the neuron in the top layer. 
The concentration of the output species becomes equal to the neural network output $f^{\ast}(x)$, which in turn is close to the desired output $f(x)$ (blue dashed line). This result is obtained for any value of the input variable $x$ and any initial concentration of the species. 
} \label{fig:2}}
\end{figure}

\begin{figure}
\begin{center}
\includegraphics[type=pdf,ext=.pdf,read=.pdf,width=175mm]{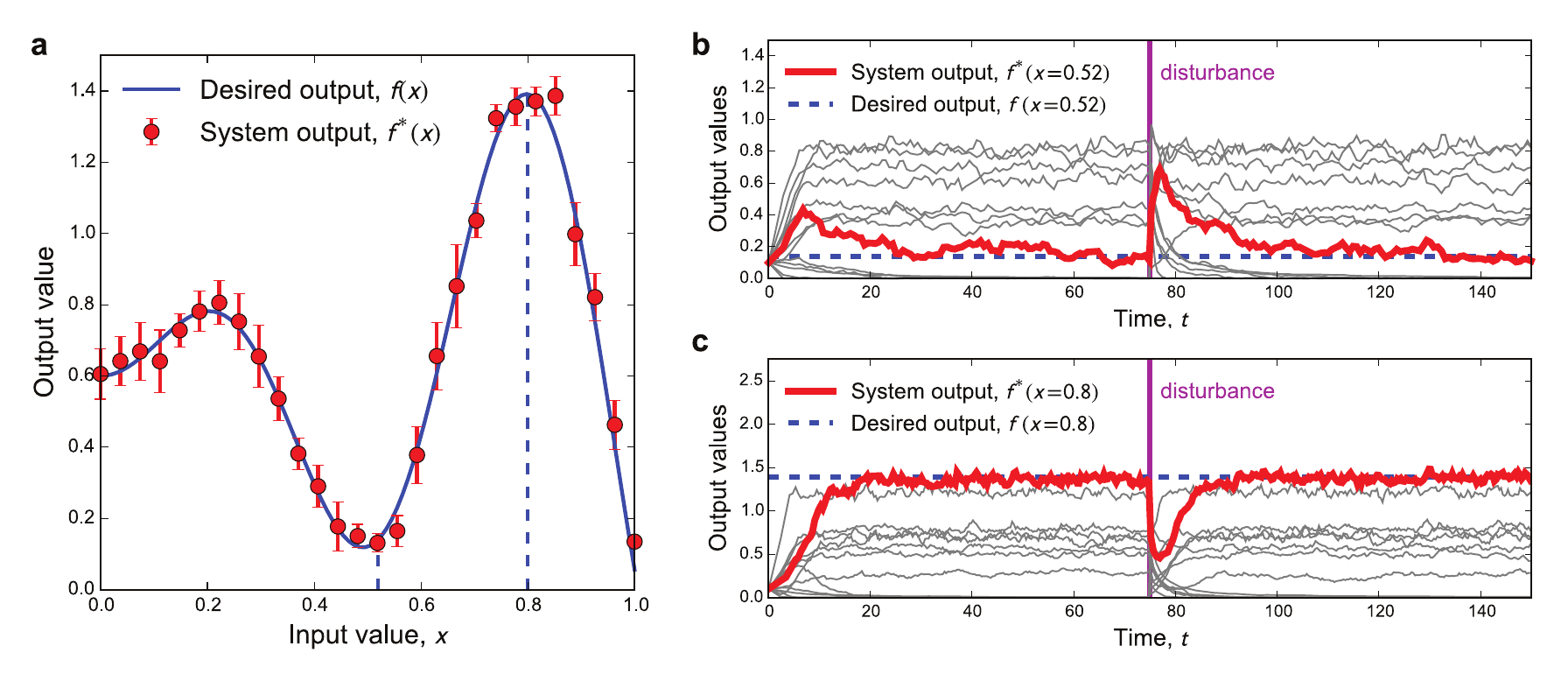}
\end{center}
\caption{{
a) Results of the stochastic simulations of the system of reactions specified in Eqs.~\ref{eq:a}, \ref{eq:ap}, \ref{eq:am}, \ref{eq:wp} and \ref{eq:wm}, where reaction rates are obtained from the weights $w$ and biases $b$ of the artificial neural network of Fig.~\ref{fig:1}a. 
For several values of the input, $x$, a stochastic simulation is performed starting from a random initial concentration of all species. 
Each red circle (and error bar) denotes the mean (and standard deviation) of the concentration of the output species at stationarity: the concentration of the output species oscillates around the desired value for all $x$. 
b) and c) Response of the system to disturbances. Two realisations of stochastic simulations corresponding to the input values $x = 0.52$ (b) and $x = 0.8$ (c). In both cases all initial concentrations are equal to $0.1$ and in the first part of the simulations (up to time $t = 75$) the system is left free to evolve to stationarity where the concentration of the output species (red line) oscillates around the desired output (blue dashed line). 
At time $t = 75$ the system is disturbed by changing the concentrations of all species to a random value. After a short transient, the system recovers from the disturbance and returns to the stationary state in which the output species oscillates around the desired output. } \label{fig:3}}
\end{figure}

Systems capable of performing different functions and tasks triggered by changes in environmental conditions or the presence of external stimuli are common in nature. 
The cell is the fundamental building block of living organisms and represents a key example of such systems: 
it is composed by a large number of interacting biological components that participate in multiple processes and reactions contributing to the system's sustenance, reproduction and function \cite{RefWorks:380}. 
Moreover, the cell is an open system in communication with the environment and it is able to sense, respond, and adapt to changing external conditions. 
Other examples include the immune system \cite{RefWorks:372,RefWorks:375}, ecological communities 
\cite{RefWorks:328,RefWorks:357,RefWorks:119}, 
social and economic systems \cite{RefWorks:46,RefWorks:155,RefWorks:363,RefWorks:362}.
From a mathematical perspective, these are complex systems out of equilibrium with many degrees of freedom coupled in a non linear way and characterised by the emergence of order and collective behaviour \cite{RefWorks:249,RefWorks:377,RefWorks:378}. 
A typical feature of many biological and ecological systems is their capability to be highly sensitive and responsive to small changes of the values of specific key variables, while being at the same time extremely resilient to a large class of disturbances \cite{RefWorks:379,RefWorks:369}. 
The possibility to build artificial systems with these characteristics is of extreme importance for the development of nanomachines and biological circuits with potential medical and environmental applications \cite{RefWorks:367,RefWorks:376,RefWorks:383}.
The main theoretical difficulty toward the realisation of these devices lies in the lack of a mathematical methodology to design the blueprint of a self-controlled system composed of a large number of microscopic interacting constituents that should operate in a prescribed fashion \cite{RefWorks:368,RefWorks:366,RefWorks:365,RefWorks:374,RefWorks:373}.
In this paper I introduce a general methodology to engineer a system of particles belonging to different species and interacting via autocatalytic reactions, 
which can automatically adjust their concentrations in order to reach a desired output configuration in response to a particular change of some input variable. 
The methodology is based on the mathematical equivalence between a particular set of autocatalytic reactions and artificial neural networks. 
An artificial neural network is made of many interacting artificial neurons organised in a series of layers \cite{RefWorks:371}. 
Each neuron of layer $l$ receives inputs from the neurons of layer $l-1$, performs a linear combination of this input and produces as output a non linear function, $a$, of this linear combination:  
\begin{align} \label{eq:neuron}
 x^{(l)}_i = a \left( \sum_j w^{(l-1)}_{ij} x^{(l-1)}_j + b^{(l-1)}_i  \right)  \, \text{,}
\end{align}
where $x^{(l)}_i$ is the output of neuron $i$ of layer $l$ and the parameters $w$ and $b$ are called weights and biases, respectively.
The nonlinear function $a(z)$ is called activation function and it is usually chosen among the logistic function, $a(z) = 1/(1+ e^{-z})$, 
the hyperbolic tangent $a(z) = \tanh(z)$, and  
the rectifier  $a(z) = \max(0,z)$.  
The first (bottom) layer of the network receives input from external variables, while the last (top) layer corresponds to the output of the network. 
Artificial neural networks can be efficiently trained to perform classification tasks and to approximate complex functions (regression). 
In supervised learning, optimisation algorithms are used to find the values of the network parameters (weights and biases) that minimise some loss function that measures the distance between the network's output and the desired output over all input values. 
At the heart of the proposed methodology is the mathematical equivalence between artificial neurons and particles interacting via autocatalytic reactions: 
each artificial neuron is identified with a species of particles and the neuron's output corresponds to the concentration, i.e. number of particles per unit volume, of that species. 
Similarly to artificial neurons in a neural network, species can be divided into a series of layers such that the concentration (or number of particles) of a species in layer $l$ is determined only by the concentrations of species in layer $l-1$. 
Specifically, each species $i$ interacts with two sets of species, input and output species: 
the concentrations of species $i$'s input species determine species $i$'s growth or decrease rate, 
while the concentration of species $i$ determines the growth or decrease rates of species $i$'s output species.
The simplest example, shown in Fig.~\ref{fig:1}a, is an artificial neural network with one neuron, which takes $n$ input signals and produces one output value. 
This corresponds to one species, $i$, characterised by the following reactions (see Fig.~\ref{fig:1}b): 
\begin{align}
2 Y_i  &\overset{1}{\to} Y_i \label{eq:a}\\
Y_i &\overset{|b_{i}|}{\to} 2 Y_i         &\quad &\text{if } b_{i} >0 &\, \label{eq:ap} \\
Y_i &\overset{|b_{i}|}{\to} \varnothing   &\quad &\text{if } b_{i} <0 \,\text{,} &\, \label{eq:am}
\end{align}
where $Y_i$ denotes one particle of species $i$, and $b_i$ is a self-interaction rate. 
The interactions of species $i$ with the $j = 1,\dots,n$ input species are specified by the reaction rates $w_{ij}$: 
\begin{align}
Y_i + X_j &\overset{|w_{ij}|}{\to} 2 Y_i + X_j   &\quad &\text{if } w_{ij} >0 &\, \label{eq:wp} \\
Y_i + X_j &\overset{|w_{ij}|}{\to} X_j           &\quad &\text{if } w_{ij} <0 \, \text{.} &\, \label{eq:wm}
\end{align}
Input particles $X_j$ are not consumed or produced in reactions \ref{eq:wp} and \ref{eq:wm}, but they collectively act as activators (Eq.~\ref{eq:wp}) or inhibitors (Eq.~\ref{eq:wm}) of the production of $Y_i$ particles without changing the number of particles of the input species. 
The set of reactions in Eqs. \ref{eq:a}, \ref{eq:ap}, \ref{eq:am}, \ref{eq:wp} and \ref{eq:wm}, corresponds to the following rate equation 
\begin{align}
\frac{d}{dt} y_i =  \left( \sum_{j=1}^n w_{ij} x_j + b_i \right) y_i -  y_i^2 =  r_i y_i -  y_i^2 \label{eq:req}
\end{align}
where $y_i$ and $x_j$ indicate the concentration of particles of species $i$ (output species) and $j$ (input species) respectively. 
These rate equations correspond to a predator-prey system with a Holling type I functional response frequently used to model population dynamics of ecological systems \cite{RefWorks:381,RefWorks:262}.
The solution of Eq.~\ref{eq:req} is 
\begin{align}
y_i (t) = \frac{r_i}{1 + \gamma_0 \, e^{- r_i t}} \text{,} \label{eq:reqsol}
\end{align}
with $ r_i = \left( \sum_{j=1}^n w_{ij} x_j + b_i \right)$ and $\gamma_0$ that depends on the number of particles of species $i$ at $t=0$, $y_i(0)$.
In the long time limit the concentration of particles of species $i$, $y_i^{\infty}$, has the following simple expression 
\begin{align} \label{eq:reqstat}
y_i^{\infty}  = 
\begin{cases}
r_i  & \text{if } r_i  > 0 \\
0 & \text{if } r_i \leq 0
\end{cases}
\end{align}
which is the rectifier activation function of artificial neurons, $a(z) = \max(0,z)$. 
Equation~\ref{eq:reqstat} indicates that the production (active state) or elimination (inactive state) of $Y_i$ particles is triggered by the higher or lower concentration of particles $i$'s input species. 
In this simple example, it is possible to tune the reaction rates $w$ and $b$ such that the number of particles of species $i$ is any linear combination of the input values $x_j$. 
The output of a generic artificial neural network with several neurons and layers can be reproduced with the introduction of an equal number of species and interactions. 
A fundamental result in artificial intelligence ensures that an artificial neural network with a sufficiently large number of neurons and layers can approximate any function arbitrarily well \cite{RefWorks:370}. 
Assume for example that the desired output of the system of interest is to reproduce the nonlinear function $f(x) = x \sin(10 x) + 0.6$ for any value of the input $x$ between 0 and 1 (blue curve in Fig.~\ref{fig:2}b). 
The artificial neural network of Fig.~\ref{fig:2}a composed by three layers with 4 neurons and one output layer with a single neuron, all using the rectifier activation function, can be trained to provide a good approximation to the desired output function $f(x)$ (red circles in Fig.~\ref{fig:2}b).
The equivalent system of reactions can then be obtained following the procedure outlined above. 
First, a number of species equal to the number of neurons, 13 in this case, is considered. 
Second, the initial concentrations of the species are set to some positive values.
Third, rate equation \ref{eq:req} is integrated to obtain the concentrations of species as a function of time, noticing that the reaction rates, $r_i$, of species in layer $l$ are computed using the weights $w_{ij}$ and biases $b_i$ of the trained neural network, and the concentrations $x_j$ of the species in layer $l-1$. 
The stationary value of the concentration of the species in the top layer equals the neural network's output value for any input $x$ and for any initial concentrations of the species (red circles in Fig.~\ref{fig:2}b). 
The time evolution of the concentration of the species relative to input $x=0.8$ shown in Fig.~\ref{fig:2}c demonstrates how the concentration of the species in the top layer reaches a stationary state equal to the output value of the neural network, $f^{\ast}(0.8)=1.38$, close to the desired output $f(0.8)=1.39$.
A similar result is obtained when, instead of the deterministic rate equation, the concentrations of species are modelled using the stochastic dynamic \cite{RefWorks:364} described by the reactions in Eqs.~\ref{eq:a}, \ref{eq:ap}, \ref{eq:am}, \ref{eq:wp} and \ref{eq:wm}. 
Again, the mean value of the concentration of the species in the top layer is compatible, within one standard deviation, with the desired output value for any input $x$ and for any initial concentrations of the species (red circles in Fig.~\ref{fig:3}a). 
In order to test the resilience to disturbances, the dynamic of the system after the concentrations of all species are set to random values is studied. 
Figures \ref{fig:3}b and \ref{fig:3}c show two realisations of the stochastic dynamic relative to inputs $x = 0.8$ and $x = 0.52$, respectively, and demonstrate that after the disturbance event the concentration of the output species (red line) returns to oscillate around the desired output (dashed blue line) in both cases. 
This indicates that the system is able to reach the desired output for any values of the initial concentrations of the species and despite disturbances. 
In this framework, embedded control is naturally achieved by specifying the desired system's output in response to any possible value of the input variables, %
without the need to analyse the response of all components in order to devise feedback and stabilisation loops, or the intervention of external control signals. 
The requirement that input species in Eqs.~\ref{eq:wp} and \ref{eq:wm} are not consumed can be relaxed by assuming a separation of time scales between reactions of different layers. 
Indeed, if reactions in lower layers have characteristic time scales that are much faster than reactions in the above layer, then the particles of species belonging to lower layers are immediately replenished when consumed in the slower reactions with particles of the above layer and the output does not substantially deviate from the output of the artificial neural network.
The extension of this methodology to multidimensional input and output is straightforward.
Relating concepts from artificial intelligence to   
dynamical systems,
the results presented here demonstrate the possibility to employ approaches and techniques developed in one field to the other, bringing potential advancements in both disciplines and related applications.

\bibliographystyle{naturemag}

\newpage

\newpage

\newpage

\end{document}